\documentclass[letterpaper, 10 pt, conference]{ieeeconf}  

\usepackage{graphicx}
\usepackage{multicol}
\usepackage{makecell}
\usepackage{lipsum}
\usepackage[utf8]{inputenc}
\usepackage[T1]{fontenc}
\usepackage{mathtools}

\usepackage{balance}
\usepackage{tcolorbox}

\IEEEoverridecommandlockouts                              
\usepackage{mdframed}
\usepackage{ntheorem}
\usepackage{subfig}
\usepackage{graphicx}
\usepackage{xcolor}
\usepackage{xspace}
\usepackage{balance}

\usepackage{framed}
\title{\LARGE \bf
Predicting the Programming Language of Questions and Snippets of StackOverflow Using Natural Language Processing
}

\author{Kamel Alreshedy, Dhanush Dharmaretnam, Daniel M. German, Venkatesh Srinivasan and T. Aaron Gulliver\\
Department of Computer Science,
University of Victoria\\
PO Box 1700, STN CSC, Victoria BC, Canada V8W 2Y2\\
Kamel, Dhanushd, dmg, srinivas@uvic.ca, agullive@ece.uvic.ca}
\definecolor{red}{rgb}{1,0,0}

\begin{document}

\maketitle
\thispagestyle{empty}
\pagestyle{empty}
\theoremseparator{.}
\newmdtheoremenv{theo}{RQ}
\newmdtheoremenv{res}{RQ}
\newmdtheoremenv{box1}{}

\begin{abstract} 
Stack Overflow is the most popular Q\&A website among software developers. As a platform for knowledge sharing and
  acquisition, the questions posted in Stack Overflow usually contain a code snippet. Stack Overflow relies on users
  to properly tag the programming language of a question and it simply assumes that the programming language of the snippets inside
  a question is the same as the tag of the question itself.
In this paper, we propose a classifier to predict the programming language of questions posted in Stack Overflow using Natural Language Processing (NLP) and Machine Learning (ML).
The classifier achieves an accuracy of 91.1\% in predicting the 24 most popular programming languages by combining
features from the title, body and the code snippets of the question.  We also propose a classifier that 
only uses the title and body of the question and has an accuracy of 81.1\%.
Finally, we propose a classifier of code snippets only that achieves an accuracy of 77.7\%.
These results show that deploying Machine Learning techniques on the combination of text and the code
snippets of a question provides the best performance. These results demonstrate also that it is possible to identify
the programming language of a snippet of few lines of source code. We visualize the feature space of two programming languages `Java and SQL' in order to identify some special properties of information inside the questions in Stack Overflow corresponding to these languages.  

\end{abstract}

 
\section{INTRODUCTION}
 
In the last decade, Stack Overflow has become a widely used resource in software development.
Today inexperienced programmers rely on Stack Overflow to address questions they have regarding their software
development activities.

Along with the growth of Stack Overflow, the number of programming languages in use has increased.  In its 2018
Developer's Survey, Stack Overflow lists
38 different programming languages in its list of most ``loved'', ``dreaded'' and ``wanted'' languages.  The TIOBE Programming Language index
tracks more than 100 languages \cite{c32}. 


Forums like Stack Overflow rely on the tags of questions to match them to users who can provide answers.  However,
new users in Stack Overflow or novice developers may not tag their posts correctly.  This leads to posts being downvoted and flagged by moderators even though the question may be relevant and adds value to the community.  In some
cases, Stack Overflow questions that are related to programming languages may lack a programming language tag. For
example, Pandas is a popular Python library that provides data structures and powerful data analysis tools; however, its
Stack Overflow questions usually do not include a Python tag.
This could create confusion among developers who may be new to
a programming language and might not be familiar with all of its popular libraries. The problem of missing language tags
could be addressed if posts are automatically tagged with their associated programming languages.
 
Another related problem is the identification of the programming language of a snippet of code. Given a few lines of
code, it is often necessary to identify the language in which they are written. Stack Overflow relies on
the tag of a question to determine how to typeset any snippet in it. If a question is not tagged
with any programming language, then the code is not typeset; however, the moment a tag is added, the code is rendered with
different colours for the different syntactic constructs of the language. If a question has snippets in two or more
languages, Stack Overflow will only use the first programming language tag to typeset all the snippets of the question.

The problem of identifying the language of snippets spans beyond Stack Overflow. Snippets are widely included in blog
posts and stored in cut-and-paste tools (such as Github's Gists and Pastebin). They might also be stored locally by the
user in tools such as QSnippets or Dash.
Gists require the user to give each snippet a
filename, which it uses to classify the programming language. Pastebin, like most snippet management tools, requires the user to select the language of the snippet manually. In both cases, the onus is on the developer to specify the language. 
Most tools that use the source code in documentation (such as recommenders) typically require that the document is already
tagged with a programming language to be processed (and assumes all the snippets in it are written in that language).

In this paper, we evaluate the use of Machine Learning (ML) models to predict the programming languages in Stack
Overflow questions. Our research questions are:

\begin{enumerate}
\item \textbf{RQ1. Can we predict the programming language of a question in Stack Overflow?}
\item \textbf{RQ2. Can we predict the programming language of a question in Stack Overflow without using
    code snippets inside it?}
\item \textbf{RQ3. Can we predict the programming language of code snippets in Stack Overflow questions?}  
\end{enumerate}
 
For the first research question, we are interested in evaluating how machine learning performs while trying to identity the language of a question when
  all the information in a Stack Overflow question is used; this includes its title, body (textual information) and code snippets in it.
   The purpose of the second research question is to determine if the inclusion of code snippets is an essential factor to determine the programming language that a question refers to.
    And finally, the purpose of the third research question is to evaluate the ability to use machine learning to predict the language of a snippet of source code; a successful predictor will have applications beyond Stack Overflow, as it could also be applied to snippet management tools and code search engines that scan documentation and blogs for relevant information.

The main contributions of the paper are as follows:



\begin{enumerate}
\item
A prediction method that uses a combination of code snippets and textual information in a Stack Overflow question.
This classifiers achieves an accuracy of 91.1\%, precision is 0.91 and recall 0.91  in predicting the tag of programming language.
\item
A classifier that uses only textual information in Stack Overflow questions to predict their programming language.
This classifier achieves an accuracy of 81.1\%, a precision of 0.83 and a recall of 0.81  which is much higher than the previous best model (Baquero \textit{et al.}~\cite{c18}).
\item
A prediction model based on Random Forest \cite{c34} and XGBoost \cite{c33} classifiers that predicts the programming language using only a code snippet in a Stack Overflow question.
This model is shown to provide an accuracy of 77.7\%, a precision of 0.79 and a recall of 0.77.
\item 
Use of Word2Vec \cite{c20} to study the features of two programming languages, Java and SQL; these features are projected into a $300$ dimensional vector space using Word2Vec and visualized using t-SNE.
\end{enumerate}

The rest of the paper is organized as follows. We begin by describing the dataset extraction and processing in Section~\ref{sec:DatasetExtractionandProcessing}. Our methodology is described in Section~\ref{sec:methodology}. The results and discussion are presented in Section~\ref{sec:randd} and Section~\ref{sec:discussion} respectively. Sections~\ref{sec:relatedwork} and \ref{sec:futurework} discuss related and future work. Finally, threats to validity and conclusions are outlined in the last two sections of the paper (Sections \ref{sec:threatstovalidity} and \ref{sec:conclusion}).

\section{Dataset Extraction and Processing}\label{sec:DatasetExtractionandProcessing}
 
In this section, the details of the Stack Overflow dataset are discussed.
Then, the preprocessing steps used for data extraction and processing are explained.

\subsection{Stack Overflow Selection}


As of July 2017, Stack Overflow had 37.21 million posts, of which 14.45 million are questions with 50.9k different tags.
In this paper, the programming language tags in Stack Overflow are of interest.
The most popular 24 programming languages as per the 2017 Stack Overflow developer survey were selected for analysis~\cite{c22}.
They constitute about 93\% of the questions in Stack Overflow.
The languages selected for our study are:
Assembly, C, C\#, C++, CoffeeScript, Go, Groovy, Haskell, Java,  JavaScript, Lua, Matlab, Objective-c, Perl, PHP, Python, R, Ruby, Scala, SQL, Swift, TypeScript, Vb.Net, Vba.



\subsection{Extraction and Processing of Stack Overflow Questions}

\begin{figure}[htbp]
    \centering
    \subfloat[Before applying NLP techniques.] 
  { \includegraphics[width=9cm]{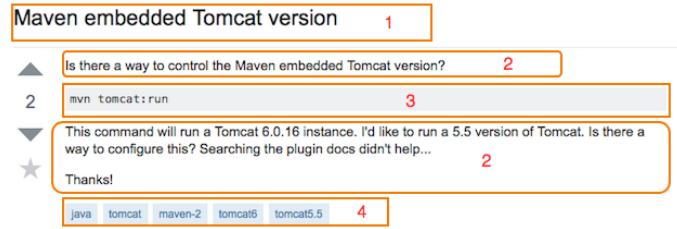}
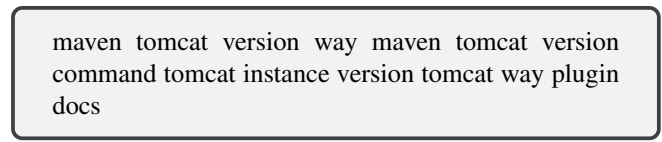
    }%
    \qquad    
    \subfloat[After applying NLP techniques.] {\begin{tcolorbox}
maven tomcat version way maven tomcat version command tomcat instance version tomcat way plugin docs
\end{tcolorbox}}%
    \caption{An example of a Stack Overflow Question.}
    \label{fig:ExtractionDataset}%
\end{figure}

The Stack Overflow July 2017 data dump was used for analysis. 
In our study, questions with more than one programming language tag were
removed to avoid potential problems during training. 
Questions chosen contained at least one code snippet, and the code snippet had at least $10$ characters.
For each programming languages $10,000$ random questions were extracted; however, two programming languages had less
than $10,000$ questions: Coffee Script (4,267) and Lua (8,460). The total number of
questions selected was 232,727. Fig. \ref{fig:ExtractionDataset} (a) shows an example of Stack Overflow post \footnote{https://stackoverflow.com/questions/1642697/}. It contains (1) the title of the post (2) the text body (3) the code snippet and (4) the tags of the post. It should be noted that the tags of the question were removed and not included as a part of text features during the training process to eliminate any inherent bias.

\begin{figure}[h]
	\centering
\frame{ \includegraphics[width=8cm,height=15cm]{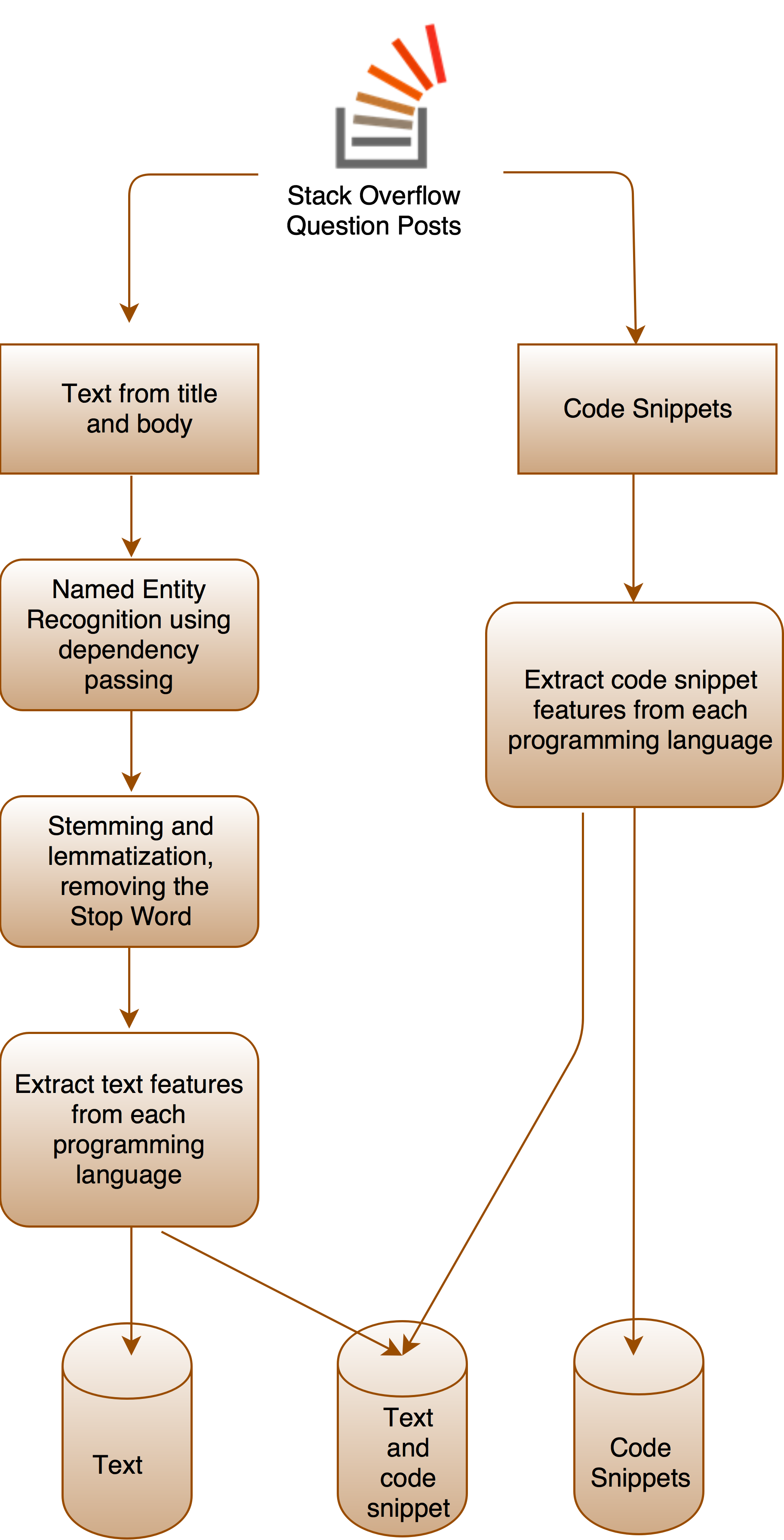}}
  \caption{The dataset extraction process.}
  \label{fig:ProcDiagram}
\end{figure}

The .xml data was parsed using xmltodict and the Python Beautiful Soup library to extract the code snippets and text from each question separately. See Fig \ref{fig:ProcDiagram}.
A Stack Overflow question consists of a title, body  and code snippet. In some cases, a question contained multiple code snippets; these were combined into one. The questions were divided into three datasets: their titles, their bodies and their snippets. 

The title and body (to which we refer to as textual information) and code snippet were used to answer the first
research question. The textual information was used to answer the second research question. Finally, the code
snippets were used to answer the last research question.


Machine learning models cannot be trained on raw text because their performance is affected by noise present in the data. The textual information (title and body) need to be cleaned and prepared before the machine learning model can be trained to provide a better prediction. Few preprocessing steps were required to clean the text. First, the non-alphanumeric characters such as punctuation, numbers and symbols were removed. Second, the entity names were identified using the dependency parsing of the Spacy Library ~\cite{c23}. An entity name is proper noun (for example, the name of an organization, company, library, function etc.). Third, the stop words such as \textit{after, about, all, and from etc.} were removed. Fourth, since the entity name can have different forms (such as study, studies, studied and studious), it is useful to train using one of these words and predict the text containing any of the word forms. To achieve this goal, stemming and lemmatization was performed using the NLTK library in Python ~\cite{c12}. At the end of all the preprocessing steps, the remaining words were used as features to help train machine learning model. Fig. \ref{fig:ExtractionDataset}(a) is the original Stack Overflow post and Fig. \ref{fig:ExtractionDataset}(b) is Stack Overflow post after application of NLP techniques.


\begin{figure*}
  \includegraphics[width=\textwidth,height=7cm]{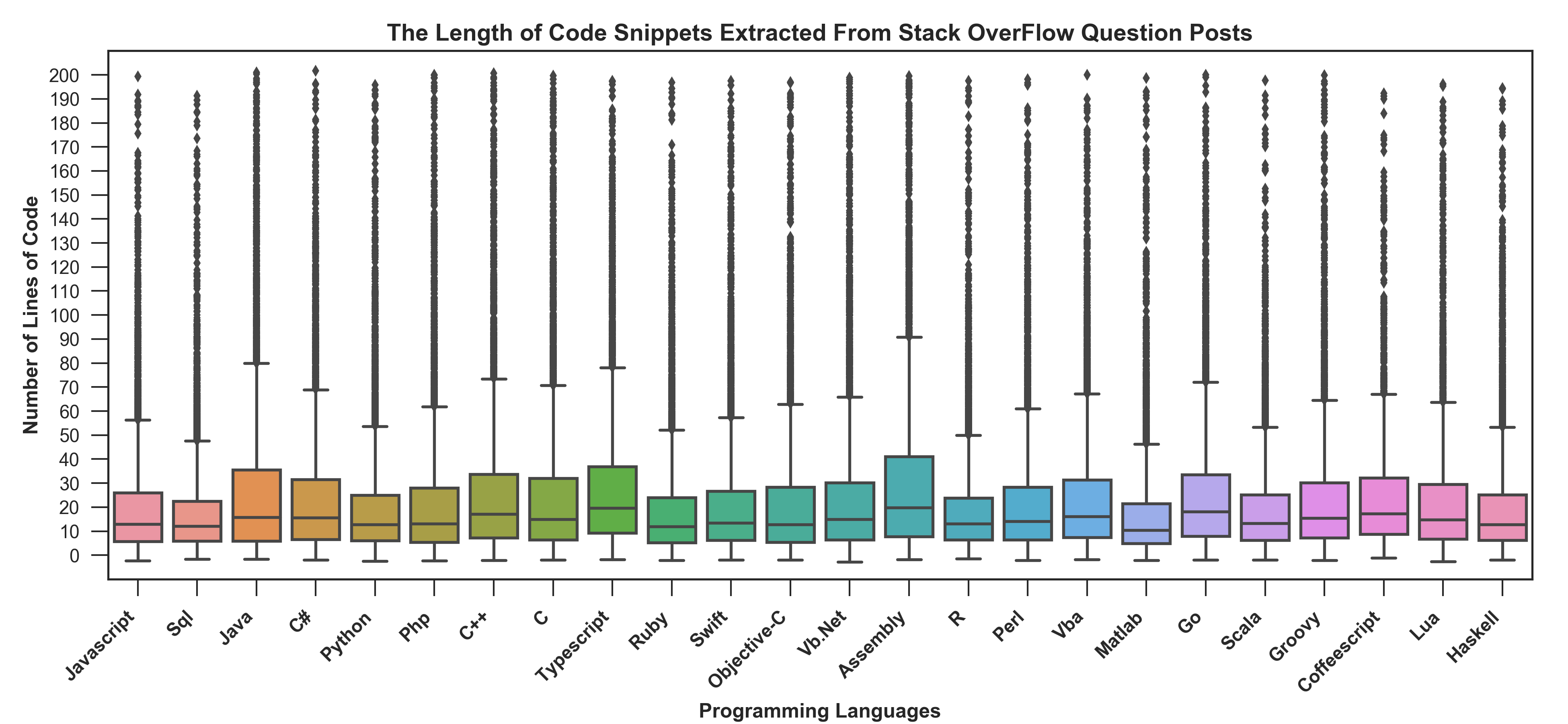}
  \caption{Box plots showing the number of lines of code in the extracted code snippets for all the languages. It should also be noted here that there were at least 400 posts which had more than 200 lines of code, however, were not included while making this plot.}
  \label{fig:BOXPLOT}
\end{figure*}


The extracted set of questions provide a good coverage of 
different versions of a programming languages. For example, code snippets were extracted for the python tags: python-3.x,
python-2.7, python-3.5, python-2.x, python-3.6, python-3.3 and python-2.6, for the Java tags java-8 and java-7, and for
the C++ tags c++11, c++03, c++98 and c++14. The snippets extracted had a significant variation in lines of code,
as shown in Fig.~\ref{fig:BOXPLOT}. The number of lines of code in the snippets varied from $6$ to $800$. 

 
\section{Methodology}\label{sec:methodology}
Textual information (title and body) and code snippets extracted from the Stack Overflow questions were split using the Term Frequency-Inverse Document Frequency (Tf-IDF) vectorizer from the Scikit-learn library \cite{c13}.
The Minimum Document Frequency (min-df) was set to 10, which means that only words present in at least ten documents
were selected (a document can be either the code snippet, the textual information, or both---code snippet and textual information).
This step eliminates infrequent words from the dataset which helps machine learning models learn from the most important vocabulary.
The Maximum Document Frequency (max-df) was set to default because the stop words were already removed in the data preprocessing step discussed in Section \ref{sec:DatasetExtractionandProcessing}.

\subsection{Classifiers}

The ML algorithms Random Forest Classifier (RFC) and XGBoost (a gradient boosting algorithm) were employed. These
algorithms provided the higher accuracy compared to the other algorithms we explored, ExtraTree and MultiNomialNB.  The
performance metrics used in this paper for the classifiers are precision, recall, accuracy, F1 score and confusion
matrix.

\subsubsection{Random Forest Classifier (RFC)}
RFC \cite{c34} is an ensemble algorithm which combines more than one classifer. This classifier generates a number of decision trees from randomly selected subsets of training dataset. Each subset provides a decision tree that votes to make the final decision during test. The final decision made depends on the decision of majority of trees. One advantage of this classifier is that if one or few of trees make a wrong decision, it will not affect the accuracy of the result significantly. Also, it avoids the overfitting problem seen in the Decision Tree model.The total number of trees in the forest is extremely important parameter because a large number of trees in the forest give high accuracy.

\subsubsection{XGBoost}
XGBoost \cite{c33}, standing for ``Extreme Gradient Boosting'', is a tree based model similar to Decision Tree and RFC. The idea behind boosting is to modify the weak learner to be a better learner. Recall that Random Forest is a simple ensemble algorithm that generates many subtrees and each tree predicts the output independently. The final output will be decided by the majority of the votes from the subtrees. However, the XGBoost is more intelligent because each subtree makes the prediction sequentially. Hence, each subtree learns from the mistakes that were made by the previous subtree. The idea of XGBoost came from gradient boosting, but XGBoost uses the regularized model to help control overfitting and give a better performance.

 

The machine learning models were tuned using RandomSearchCV, which is a tool for parameter search in the Scikit-learn library.
The XGBoost algorithm has many parameters, such as minimum child weight, max depth, L1 and L2 regularization, and evaluation metrics such as Receiver Operating Characteristic (ROC), accuracy and F1 score.
RFC is a bagging classifier and has a parameter number of estimators which is the number of subtrees used to fit the model.
It is important to tune the models by varying these parameters.
However, parameter tuning is computationally expensive using a technique such as grid search.
Therefore, a deep learning technique called Random Search (RS) tuning is used to train the models.
All model parameters were fixed after RS tuning on the cross-validation sets (stratified ten-fold cross-validation).
For this purpose, the datasets were split into training and test data using the ratio 80:20.

An important contribution of this paper is the study of the vocabulary and feature space 
for two programming languages (Java and SQL).
A word2vec model \cite{c20} was used to visualize the features from code snippets and textual information (title and body) datasets using Gensim, which is a Python framework for vector space modelling \cite{c19}.
The resulting model represented each word in the vocabulary in a $300$ dimensional vector space. The selection of $300$ as the number of dimensions for the trained word-vector is as per the recommendation of the original paper by T. Mikolov  \cite{c31}.
It is impossible to visualize concepts in such a large space, so T-SNE \cite{c21} was used to reduce the number of dimensions to $2$.
The top frequent 3\%  words were selected from the vectors for 
`Java and SQL' and analyzed using word similarity and cosine distance.
The code snippets and textual information features which are close to each other in the vector space were selected for Java and SQL using the cosine distance. Then, these features were visualized to understand the similarities and differences between Java and SQL. 


\section{Results}
\label{sec:randd}
In this section, the results obtained for the three research questions are described in detail.


\textbf{RQ1. Can we predict the programming language of a question in Stack Overflow?}\newline

\begin{figure*}[h]
  \centering
  \includegraphics[width=\textwidth]{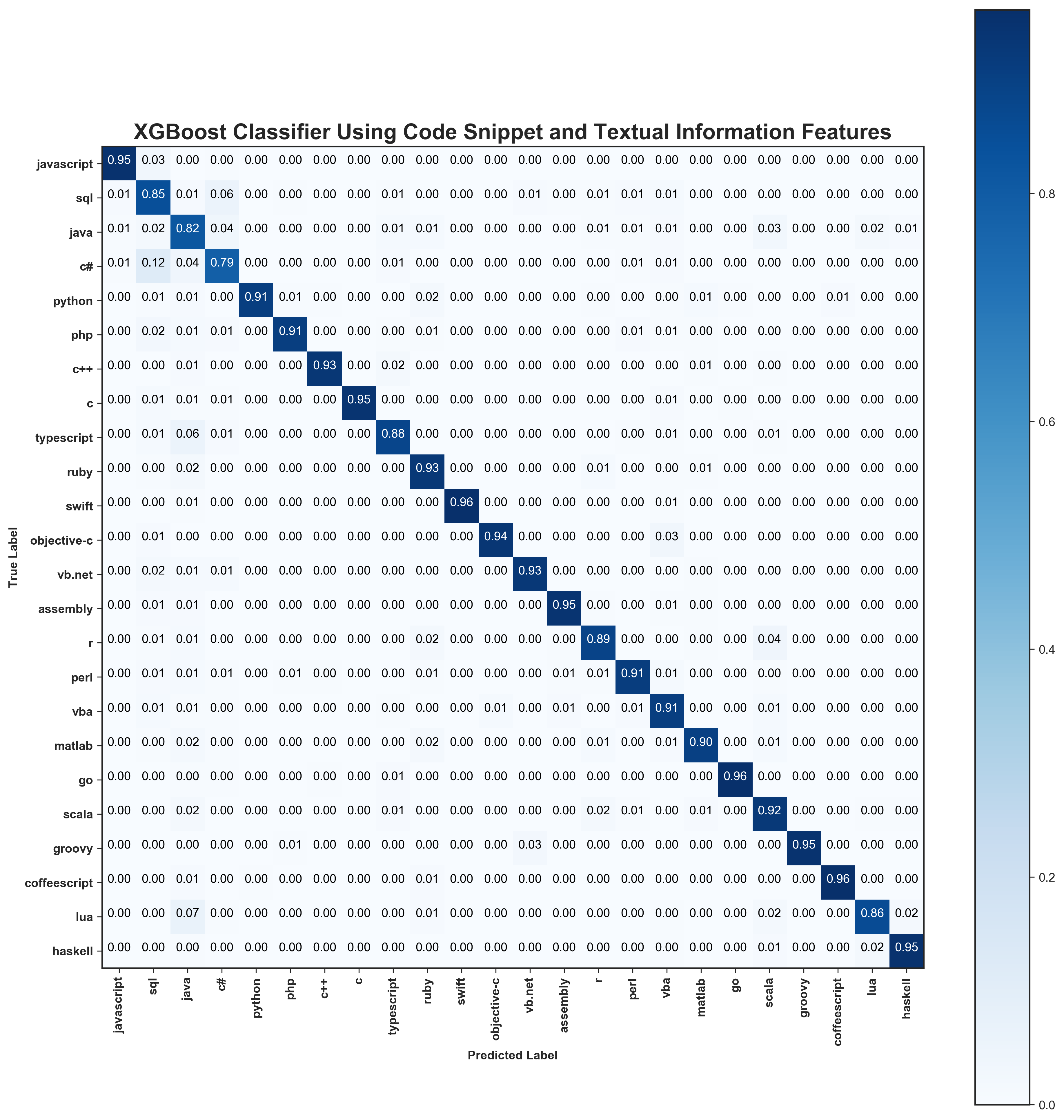}
  \caption{Confusion matrix for the XGboost classifier trained on code snippet and textual information features. The diagonal represent the percentage of programming language that was correctly predicted.}
  \label{fig:CodeText}
\end{figure*}

To answer this question, XGBoost and RFC classifiers were trained on the combination of textual information and code
snippet datasets. The XGBoost classifier achieves an accuracy of 91.1\%, and the average score for precision, recall and
F1 score are 0.91, 0.91 and 0.91 respectively; on the other hand, the FRC achieves an accuracy of 86.3\%, and the
average score for precision, recall and F1 score are 0.87, 0.86 and 0.86 respectively. The results for XGBoost
classifier are discussed in further detail because it provides the best performance. In Table \ref{Table:COdeText}, the
performance metrics for each programming language with respect to precision, recall and F1 score are given for
XGBoost. Most programming languages have a high F1 score:
 Swift (0.97), GO (0.97), Groovy (0.97) and Coffeescript (0.97) had the highest, while java (0.75), SQL (0.78), C\# (0.80) and Scala (0.88) had the lowest. 
Fig. \ref{fig:CodeText} shows the performance of the XGBoost classifier as a confusion matrix.

\begin{table}[h]
  \centering
  \begin{tabular}{| c | c | c | c |}
   \hline
Programming  & Precision & Recall & F1-score \\ \hline
Swift        & 0.98      & 0.96   & 0.97     \\ \hline
Go           & 0.98      & 0.96   & 0.97     \\ \hline
Groovy       & 0.99      & 0.95   & 0.97     \\ \hline
Coffeescript & 0.98      & 0.96   & 0.97     \\ \hline
Javascript   & 0.97      & 0.95   & 0.96     \\ \hline
C            & 0.98      & 0.95   & 0.96     \\ \hline
C++          & 0.97      & 0.93   & 0.95     \\ \hline
Objective-c  & 0.97      & 0.94   & 0.95     \\ \hline
Assembly     & 0.96      & 0.95   & 0.95     \\ \hline
Haskell      & 0.95      & 0.95   & 0.95     \\ \hline
Python       & 0.97      & 0.91   & 0.94     \\ \hline
Vb.net       & 0.95      & 0.93   & 0.94     \\ \hline
PHP          & 0.94      & 0.91   & 0.93     \\ \hline
Ruby         & 0.89      & 0.93   & 0.91     \\ \hline
Perl         & 0.91      & 0.91   & 0.91     \\ \hline
Matlab       & 0.92      & 0.90    & 0.91     \\ \hline
R            & 0.91      & 0.89   & 0.90      \\ \hline
Lua          & 0.94      & 0.86   & 0.90      \\ \hline
Typescript   & 0.90       & 0.88   & 0.89     \\ \hline
Vba          & 0.85      & 0.91   & 0.88     \\ \hline
Scala        & 0.85      & 0.92   & 0.88     \\ \hline
C\#          & 0.81      & 0.79   & 0.80      \\ \hline
CQL          & 0.73      & 0.85   & 0.78     \\ \hline
Java         & 0.70       & 0.82   & 0.75     \\ \hline
 \end{tabular}
  \caption{Performance for the proposed classifier trained on textual information and code snippet features.}
  \label{Table:COdeText}%
\end{table}


\medskip
\textbf{RQ2. Can we predict the programming language of a question in Stack Overflow without using code snippets inside it?}\newline

To answer this research question, two machine learning models were trained using the XGBoost and RFC classifiers on the
dataset that contained only the textual information.  The XGBoost classifier achieved an accuracy of 81.1\%, and the
average score for precision, recall and F1 score were 0.83, 0.81 and 0.81 respectively. In Table
\ref{Table:textualinformation}, the performance metrics for each programming language with respect to precision, recall
and F1 score are given for XGBoost. RFC achieved slightly lower performance than XGboost, 
with an average score for precision, recall and F1 score of 0.76, 0.74 and 0.75 respectively. Note that the accuracy of
XGBoost using textual information decreased by about 10\% compared to using both the textual
information and its code snippet.
The top performing languages based on the F1 score are coffeescript (0.94), javascript (0.92), swift (0.92), Go (0.92), Haskell (0.92), C (0.91) Objective-C (0.90) and Assembly (0.89). Note further that the F1 scores of most of
the programming languages in the table decreased by approximately 5\%
with a few exceptions (such as vb.net, vba, PHP, Lua, C\#, SQL and Java). It should be further noted that the languages CoffeeScript, Javascript, Swift, Go and C have a high F1 score and performed very well in both the (RQ1) and (RQ2). Java and SQL have the worst performance metrics in (RQ1) and (RQ2), especially in (RQ2) the F1 score decreased by as much as 25\%.

\begin{table}[h]
\centering

\begin{tabular}{|l|l|l|l|}
\hline
Programming  & Precision & Recall & F1-score \\ \hline
Coffeescript & 0.96      & 0.91   & 0.94     \\ \hline
Javascript   & 0.94      & 0.89   & 0.92     \\ \hline
Swift        & 0.94      & 0.89   & 0.92     \\ \hline
Go           & 0.95      & 0.89   & 0.92     \\ \hline
Haskell      & 0.92      & 0.91   & 0.92     \\ \hline
C            & 0.93      & 0.88   & 0.91     \\ \hline
Objective-c  & 0.94      & 0.87   & 0.90      \\ \hline
Assembly     & 0.92      & 0.87   & 0.89     \\ \hline
Python       & 0.95      & 0.82   & 0.88     \\ \hline
Groovy       & 0.95      & 0.82   & 0.88     \\ \hline
C++          & 0.92      & 0.83   & 0.87     \\ \hline
Ruby         & 0.86      & 0.88   & 0.87     \\ \hline
R            & 0.88      & 0.82   & 0.85     \\ \hline
Perl         & 0.88      & 0.81   & 0.84     \\ \hline
Matlab       & 0.88      & 0.80    & 0.84     \\ \hline
Scala        & 0.80       & 0.90    & 0.84     \\ \hline
Typescript   & 0.86      & 0.80    & 0.83     \\ \hline
Vb.net       & 0.82      & 0.82   & 0.82     \\ \hline
Vba          & 0.76      & 0.81   & 0.78     \\ \hline
PHP          & 0.82      & 0.72   & 0.77     \\ \hline
Lua          & 0.73      & 0.59   & 0.65     \\ \hline
C\#          & 0.65      & 0.61   & 0.63     \\ \hline
SQL          & 0.42      & 0.75   & 0.54     \\ \hline
Java         & 0.43      & 0.58   & 0.49     \\ \hline
\end{tabular}
\caption{Performance for the proposed classifier trained on textual information features.}
\label{Table:textualinformation}
\end{table}



\begin{table}[htbp]
  \centering
  \begin{tabular}{| c | c | c | c |}
\hline
Programming  & Precision & Recall & F1-score \\ \hline
Javascript   & 0.94      & 0.88   & 0.91     \\ \hline
Coffeescript & 0.92      & 0.86   & 0.89     \\ \hline
PHP          & 0.91      & 0.85   & 0.88     \\ \hline
Go           & 0.92      & 0.84   & 0.87     \\ \hline
Groovy       & 0.91      & 0.84   & 0.87     \\ \hline
Swift        & 0.91      & 0.82   & 0.86     \\ \hline
C            & 0.86      & 0.82   & 0.84     \\ \hline
Vb.net       & 0.89      & 0.81   & 0.84     \\ \hline
Haskell      & 0.87      & 0.8    & 0.83     \\ \hline
C++          & 0.86      & 0.78   & 0.82     \\ \hline
Vba          & 0.82      & 0.77   & 0.80      \\ \hline
Lua          & 0.87      & 0.74   & 0.80      \\ \hline
Assembly     & 0.76      & 0.76   & 0.76     \\ \hline
Python       & 0.85      & 0.67   & 0.75     \\ \hline
Ruby         & 0.72      & 0.79   & 0.75     \\ \hline
Matlab       & 0.79      & 0.72   & 0.75     \\ \hline
Scala        & 0.71      & 0.79   & 0.75     \\ \hline
SQL          & 0.70       & 0.77   & 0.73     \\ \hline
C\#          & 0.78      & 0.68   & 0.73     \\ \hline
Perl         & 0.75      & 0.72   & 0.73     \\ \hline
R            & 0.72      & 0.70    & 0.71     \\ \hline
Typescript   & 0.68      & 0.66   & 0.67     \\ \hline
Java         & 0.64      & 0.69   & 0.66     \\ \hline
Objective-c  & 0.42      & 0.85   & 0.56     \\ \hline
\end{tabular}
\caption{Performance for the proposed classifier trained on code snippet features.}
\label{Table:Code}
\end{table}

\begin{table*}[htbp]
  \centering
  \vspace{8pt}
  \setlength\tabcolsep{4pt}
  \begin{tabular}{| l | l | r | r | r | r |}
     \hline
    Model & Description & Accuracy & Precision & Recall & F1 score \\ \hline
    \textbf{Previous} &&&&&\\
    \hline
    \quad Baquero [18] code snippets & \makecell{A model trained using Support Vector Machine on \\question questions from Stack Overflow using code features}  & 44.6\% & 0.45 & 0.44  & 0.44 \\ \hline
    \quad Baquero [18] textual information & \makecell{A model trained using Support Vector Machine on \\ questions from Stack Overflow using text features}&  60.8\% & 0.68 &  0.60 & 0.60\\ \hline
    \textbf{Proposed}&&&&&\\
    \hline
    \quad code snippet features &\makecell {XGBoost classifier trained on Stack Overflow \\ questions using code snippet features}  & 77.7\% & 0.79 & 0.77  & 0.78\\ \hline
    \quad textual information  features &\makecell {XGBoost classifier trained on Stack Overflow \\questions using textual information features}  & 81.1\% & 0.83 & 0.81& 0.81\\ \hline
    \quad code snippet and textual information features &\makecell {XGBoost classifier trained on Stack Overflow questions\\ using code snippets and textual information features}  & 91.1\%  & 0.91 &  0.91 & 0.91\\
    \hline
  \end{tabular}
  \caption{A Comparison of Previous and Proposed classifiers}
  \label{Table:Summary}
\end{table*}

\begin{figure*}[h]
  \centering
  \includegraphics[width=18cm]{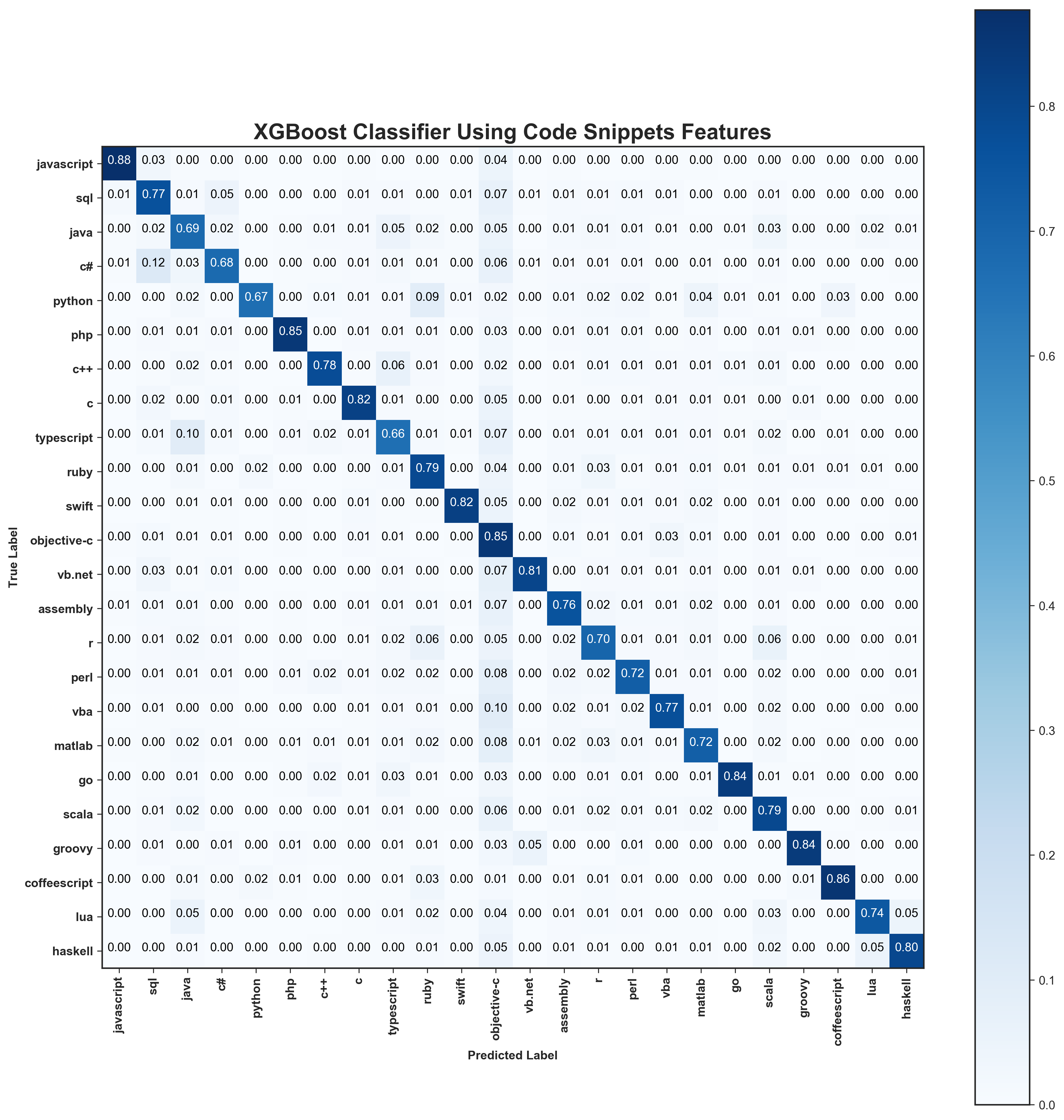}
  \caption{Confusion matrix for the XGboost classifier trained on code snippet features. The diagonal represent the percentage of programming language that was correctly predicted.}
  \label{fig:CodeConf}
\end{figure*}

\medskip
\textbf{RQ3. Can we predict the programming language of code snippets in Stack Overflow questions?}\newline
To predict the programming language from a given code snippet, two ML classifiers were trained on the code snippet
dataset. XGBoost achieved an accuracy of 77.7\%, and the average score for precision, recall and F1 score are 0.79, 0.77 and 0.77 respectively. In Table \ref{Table:Code}, the performance metrics for each programming language with respect to precision, recall and F1 score are given while using XGBoost. RFC achieved accuracies of accuracy of 70.1\%, and the average score for precision, recall and F1 score are 0.72, 0.72 and 0.70 respectively. The results obtained for the code snippet dataset show the worst performance for both classifiers.
The programming languages JavaScript (0.91), CoffeeScript (0.89) and PHP (.88) had a good F1 score. 
The F1 score of PHP in (RQ1) is close to the average and in (RQ2) is one of the worst; however, in (RQ3) the PHP language has the third highest F1 score. Objective-C has the worst F1 score and precision (0.56 and 0.42); but its recall is extremely high (0.85). When the programming language of a code snippet is extremely hard to identity, XGBoost frequently misclassified it as Objective-C, while we observed that RFC misclassified such snippets as Typescript. We have looked manually at some of these code snippets and are not able to identify the programming language easily\footnote{https://stackoverflow.com/questions/855360/}\footnote{https://stackoverflow.com/questions/942772/}\footnote{https://stackoverflow.com/questions/9986404/}\footnote{https://stackoverflow.com/questions/2115227/}. This is the main motivation for combining the textual information and code snippet in (RQ1). If the classifier gets confused when predicting the programming langauges from a code snippet, the textual information (title and body) will help the machine learning model to make a better prediction. Fig. \ref{fig:CodeConf} shows the confusion matrix for the XGBoost classifier. Table \ref{minchara} shows how the accuracy improves as the minimum size of the code snippet in the dataset is increased from 10 to 100 characters.

The comparison between the results of textual information dataset and code snippet dataset shows a slightly difference in
accuracy of 5\% (in average). On the other hand, using the combination of textual information and code snippet significantly increased the accuracy by 10\% compared to using only the textual information and 14\% compared to using only code snippets. Since many Stack Overflow posts can have a large textual information and a small code snippet or vice versa, combining the two gives a high accuracy in (RQ1). 


\section{Discussion}\label{sec:discussion}

The most important observation in the previous section is that for the research question (RQ1), XGBoost achieves high accuracy of 91.1\%, while for (RQ2) and (RQ3), it only achieves an accuracy of 81.1\% and 77.7\% respectively. This observations highlights the importance of using the combining of textual information and code snippets in predicting tags in comparison to textual information or code snippet only. 
In some cases, Stack Overflow posts contain very small code snippet making it extremely hard to identify its language as many programming languages sharing the same syntax.

\begin{table}[!htb]
\centering

\begin{tabular}{|l|l|l|l|l|}
\hline
The Minimum Characters  &  Accuracy    & Precision & Recall & F1-score \\ \hline
More than 10   & 77.7\%      & 0.79   & 0.77   & 0.78   \\ \hline
More than 25   & 79.1\%      & 0.80   & 0.97   & 0.79   \\ \hline
More than 50   & 81.7\%      & 0.82   & 0.81   & 0.81   \\ \hline
More than 75   & 83.1\%      & 0.83   & 0.83   & 0.83   \\ \hline
More than 100 & 84.7\%      & 0.85   & 0.84    & 0.84  \\ \hline
\end{tabular}
\caption{Effect of the minimum number of characters in code snippet on accuracy}
\label{minchara}
\end{table}

Dependency parsing and extracting of entity names using a Neural Network (NN) through Spacy appeared to help reduce
noise and to extract important features from Stack Overflow questions. This is likely
the main reason for the significant improvement in performance compared to previous approaches in the literature.

The analysis of the feature space of the top performing languages indicates that these languages have unique code snippet features (keywords/identifiers) and textual information features (libraries, functions).
For example, when the textual information based features were visualized for Haskell, words such as `GHC', `GHCI', `Yesod' and `Monad' were obtained.
`GHC' and `GHCI' are compilers for Haskell, `Yesod' is a web-based framework and `Monad' is a functional programming paradigm (Haskell is a functional programming language).
Most of the top performing languages have a small feature space (vocabulary) as compared to more popular languages such
as Java, Vba and C\# which have a large number of libraries and standard functions, and support multiple programming
paradigms resulting in a large feature space. A large feature space adds more complexity to the ML models.

\begin{figure*}[htbp]
    \centering
    \subfloat[Java code snippet features.]{{\includegraphics[width=16cm]{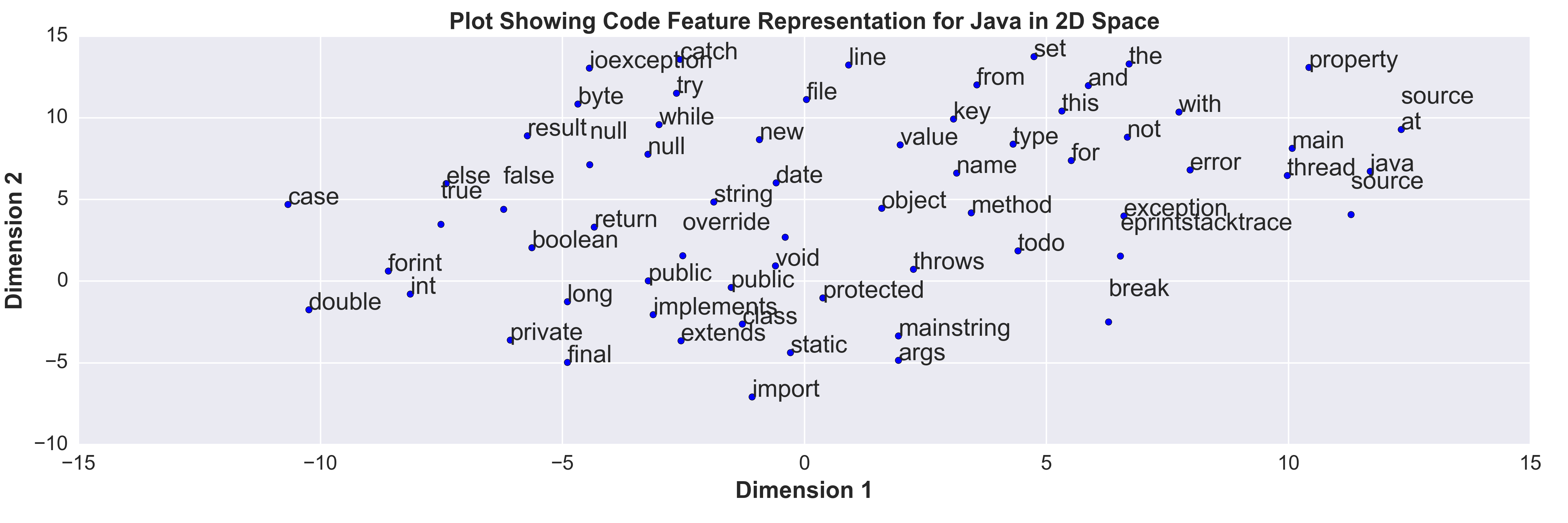} }}%
    \qquad
    \subfloat[Java textual information features.]{{\includegraphics[width=16cm]{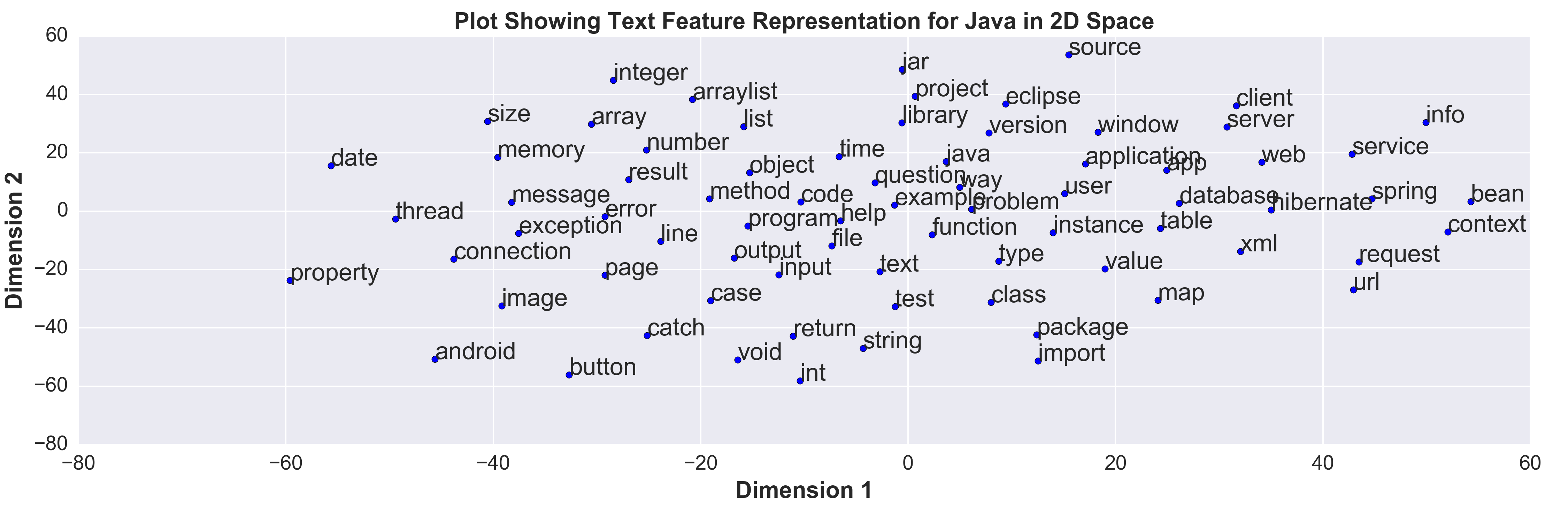} }}%
    \caption{Code snippet and textual information features of Java represented in two dimensions after using t-SNE on a trained Word2Vec model.}%
    \label{fig:Java}%
\end{figure*}

\begin{figure*}[htbp]
    \centering
    \subfloat[SQL code snippet features.]{{\includegraphics[width=16cm]{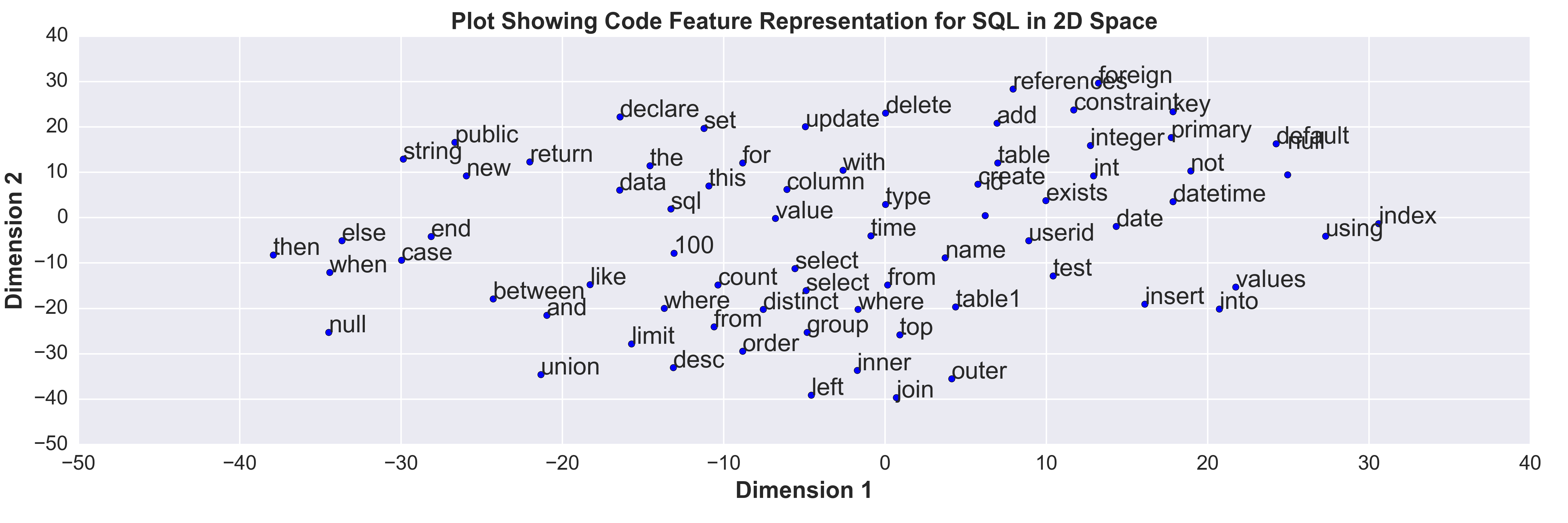} }}%
    \qquad
    \subfloat[SQL textual information features.]{{\includegraphics[width=16cm]{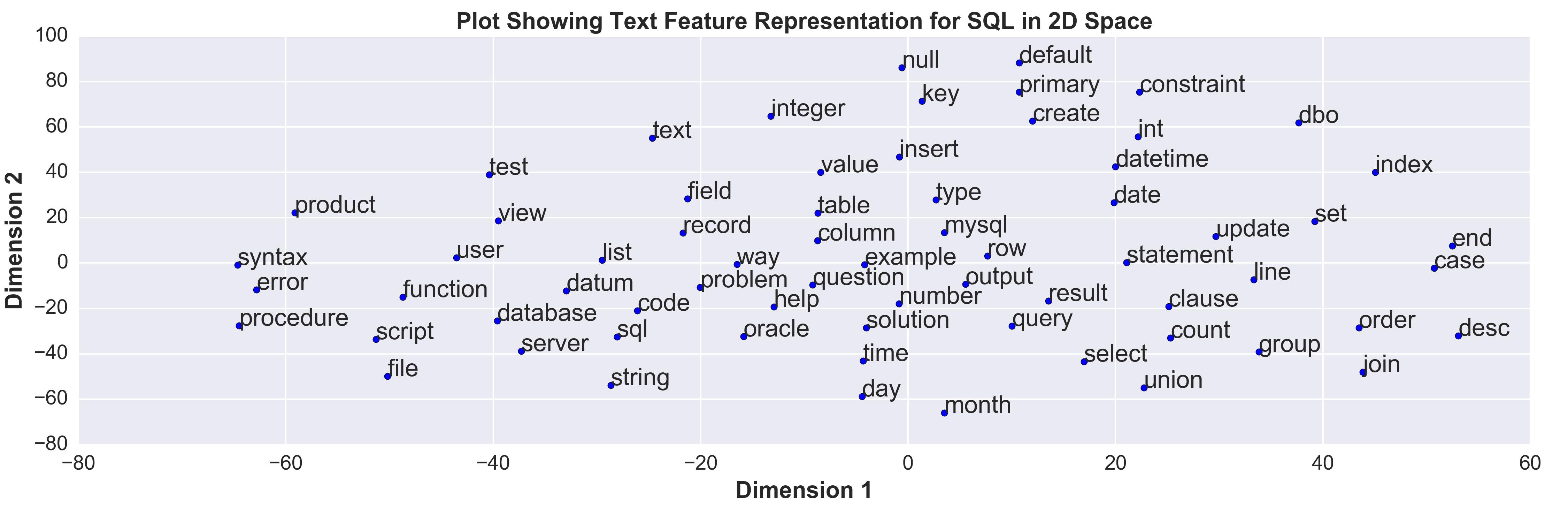} }}%
    \caption{Code snippet and text information features of SQL represented in two dimensions using t-SNE on a trained Word2Vec model.}%
    \label{fig:SQL}%
\end{figure*}

The Word2Vec models were trained in two different datasets, the code snippets and textual information, to visualize their features. Fig.~\ref{fig:Java} shows the features for Java and  Fig.~\ref{fig:SQL} shows the features for SQL.
There were more than 200 features extracted using Word2Vec for each language, but for clarity, only about 70 features are shown.
In the vector space, features which are used in the same context are close.
For example, in the Java code snippets (Fig.~\ref{fig:Java}), `public',`class',`implements' and `extends' are all close to one another in the vector space since they are typically used together in a line of code.
Further, Fig.~\ref{fig:SQL} shows that in SQL code snippets features such as `foreign', `primary', `key', `constraint' and `reference' are all used in the context of setting relationships between tables. 


The smaller cosine distance between features does not necessarily mean they are used in the same line of code but are used in the same context, e.g. declaring relationships in tables.
This is because, in Stack Overflow, developers describe their issues by including code based identifiers and features along with text not just in code blocks.
Another critical observation from the Java textual information features is that `Android' and `spring' (a framework for web and desktop applications) are located at opposite sides of the plot (maximum cosine distance).
This means that an Android developer rarely uses the spring framework for development.
Conversely, `spring' is located close to `web', `service', `url', `request' and `xml', which indicates that spring framework is frequently used in web development in Java.

\section{Related Work}\label{sec:relatedwork} 
 
Baquero \textit{et al.}~\cite{c18} proposed a classifier to predict the programming language of a Stack Overflow question.
They extracted a set of $18000$ questions from Stack Overflow that contained text and code snippets,
$1000$ questions for each of $18$ programming languages.
They trained two classifiers using a Support Vector Machine model on two different datasets which are text body and code snippet features. The evaluation achieved an accuracy of 60\% for text body features and 44\% for code snippet features which are much lower than the results obtained in this paper. 
Table~\ref{Table:Summary} summarizes our results in comparison to \cite{c18} as, to the best of our knowledge, it is the only previous work in the literature that tackles the problem of predicting programming language tags for Stack Overflow questions. 

Kennedy \textit{et al.} \cite{c28} studied the problem of using natural language identification to identify the programming language of entire
source code files from GitHub (rather than questions from Stack Overflow). Their classifier is based on five statistical language models from NLP and identifies $19$ programming languages and can achieve a high accuracy of 97.5\%. In our work, $24$
programming languages are predicted using small code snippets rather than source code file. Similarly, Khasnabish \textit{et al.} 
\cite{c27} proposed a model to detect $10$ programming languages using source code files. Four algorithms were used to
train and test the model using Bayesian learning techniques, i.e. NB, Bayesian Network (BN) and Multinomial Naive Bayes
(MNB). It was shown that MNB provides the highest accuracy of $93.48$\%.


Some editors such as Sublime and Atom add highlights to code based on the programming language.
However, this requires an explicit extension, e.g. .html, .css, .py.
Portfolio \cite{c6} is a search engine that supports programmers in finding functions that implement high-level requirements in query terms.
This engine does not identify the language, but it analyzes code snippets and extracts functions which can be reused.
Holmes \textit{et al.} \cite{c8} developed a tool called Strathcona that can find similar snippets of code.
 
 

Rekha \textit{et al.}~\cite{c10} proposed a hybrid auto-tagging system that suggests tags to users who create questions. When the post contains a code snippet, the system detects the programming language based on the code snippets and suggests many tags to users. Multinomial Naive Bayes (MNB) was trained and tested for the proposed classifier which achieved 72\% accuracy.
Saha~\textit{et al.}~\cite{c1} converted Stack Overflow questions into vectors, and then trained a Support Vector Machine using these vectors and suggested tags used the model obtained.
The tag prediction accuracy with this model is 68.47\%.
Although it works well for some specific tags, it is not effective with some popular tags such as Java.
Stanley and Byrne~\cite{c2} used a cognitive-inspired Bayesian probabilistic model
to choose the most suitable tag for a post.
This is the tag with the highest probability of being correct given the a priori tag probabilities.
However, this model normalizes the top for all questions, so it is unable to differentiate between a post where the top predicted tag is certain, and a post where the top predicted tag is questionable.
As a consequence, the accuracy is only 65\%.


\section{Future Work}\label{sec:futurework}

The study of programming language prediction from textual information and code snippets is still new, and much remains to be done. Most of the existing tools focus on file extensions rather than the code itself. In recent years, there has been tremendous progress made in the field of deep learning, especially for time series or sequence-based models such as Recurrent Neural Networks (RNNs) and Long Short-Term Memory (LSTM) networks.
RNN and LSTM models can be trained using source code one character at a time as input, but they can have a high computational cost.
 
NLP and ML techniques perform much better in predicting languages compared to tools that predict directly from code snippets. Stack Overflow text is somewhat unique in the sense that it captures the tone, sentiments and vocabulary of the developer community.
This vocabulary varies depending on the programming language. Therefore, it is important that the vocabulary for each programming language is captured, understood and separated. It is worth exploring if a CNN combined with Word2Vec can be used for this task. 

In the future, our model will be evaluated using programming blog posts, library documentation and bug repositories. This would help us understand how general the model is.

\section{Threats to Validity}\label{sec:threatstovalidity}

Construct Validity: 
In creating the datasets from Stack Overflow, only the most popular programming languages were extracted, and this was based solely on the programming language tag. However, some tags synonymous with languages were not included in the extraction process. For example, `SQL SERVER', `PLSQL' and `MICROSOFT SQL SERVER' are related to `SQL' but were discarded. 

Internal validity: After the datasets were extracted, dependency parsing was used to select entity names so as to include only the most relevant code snippet and text features. The use of dependency parsing can result in the loss of critical vocabulary and might affect our results. However, we manually analyzed the vocabulary before and after the dependency parsing to ensure that information related to the languages was not lost. Further, selecting additional features such as lines of code and programming paradigm could have improved our results but was not considered.
 
External validity: The focus of this paper was to obtain a classifier for predicting languages due to the lack of open source tools for this task. Stack Overflow was used in this study as the data source but other sources such as GitHub repositories were not explored. Therefore, no conclusions can be made about the results with other sources of code snippets and text on programming languages. Furthermore, some common programming languages such as Cobol and Pascal were not considered in this study.

\section{Conclusions}\label{sec:conclusion}


This work tackles the important problem of predicting programming languages from code snippets and textual information. In particular, it chose to focus on predicting the programming language of Stack Overflow questions. Our results show that training and testing the classifier by combining the textual information and code snippet achieves the highest accuracy of 91.1\%. Other experiments using either textual information or code snippets, achieve accuracies of 81.1\% and 77.7\% respectively. This implies that information from textual features is easier for a machine learning model to learn as compared to information from code snippet features. Our results also show that it is possible to identify the programming language of a snippet of few lines of source code.
We believe that our classifier could be applied in other scenarios such as code search engines and snippet management tool. 

\addtolength{\textheight}{-10cm}   
                                  


\end{document}